\documentclass[twocolumn,footinbib,aps,pra]{revtex4-1}
\usepackage{color}
\usepackage{amsfonts,amssymb,amsmath,mathrsfs}
\usepackage{amsbsy}
\usepackage{graphicx}
\usepackage{hyperref}

\begin{document}

\title{Analytically parameterized solutions for robust quantum control using smooth pulses}
\author{Utkan G\"ung\"ord\"u}
\email{utkan@umbc.edu}
\affiliation{Department of Physics, University of Maryland Baltimore County, Baltimore, MD 21250, USA}

\author{J.~P.~Kestner}
\affiliation{Department of Physics, University of Maryland Baltimore County, Baltimore, MD 21250, USA}

\begin{abstract}
Achieving high-fidelity control of quantum systems is essential for realization of a practical quantum computer. Composite pulse sequences which suppress different types of errors can be nested to suppress a wide variety of errors but the result is often not optimal, especially in the presence of constraints such as bandwidth limitations. Robust smooth pulse shaping provides flexibility, but obtaining such analytical pulse shapes is a non-trivial problem, and choosing the appropriate parameters typically requires a numerical search in a high-dimensional space. In this work, we extend a previous analytical treatment of robust smooth pulses to allow the determination of pulse parameters without numerical search. We also show that the problem can be reduced to a set of coupled ordinary differential equations which allows for a more streamlined numerical treatment.
\end{abstract}

\maketitle

\section{Introduction}
The main difficulty hampering the efforts to build a large scale, practical quantum computer is decoherence. Quantum error correction codes provide a promising path toward fault-tolerant quantum computers. However, a typical surface code requires access to quantum gates with a fidelity above 99\%, and significantly higher fidelities are desirable to reduce overhead. Achieving high gate fidelities in a noisy device requires carefully designed robust control fields.

There are a variety of ways to raise fidelities, and quantum control methods have been developed to pursue a number of desirable objectives such as time-optimal pulse sequences \cite{Glaser2015} (and the references therein), leakage suppression \cite{Rach2015}, or smooth pulse shaping in the absence of noise \cite{Machnes2018}.
Our work here, though, lies strictly within the subset of quantum control methods that seek to suppress stochastic logical errors by pulse shaping.

Robust composite pulse sequences \cite{Merrill2014e}, which generalize Hahn echo \cite{Hahn1950} and Carr-Purcell-Meiboom-Gill (CPMG) \cite{Carr1954,Meiboom1958} sequences to implement non-trivial unitaries, are effective for suppressing slow noise or calibration errors which remain constant during the gate time. Various pulse sequences have been developed to suppress either pulse length errors or off-resonance errors \cite{Merrill2014e}.
However, in some systems, such as spin qubits in silicon \cite{Veldhorst2014b,Veldhorst2014a,Zajac2017a,Watson2017,Yang2019} or GaAs \cite{Hanson2007,Zwanenburg2013}, noise is present in some combination of the two forms, which requires nesting these sequences \cite{Bando2013} or using specialized pulses \cite{Kestner2013,Wang2014a,Buterakos2018}. Such methods are often designed with square pulses in mind, although they can be modified to use smooth ramping profiles \cite{Wang2014a, Gungordu2018a}. However, the finite bandwidth of a physical control field may be more naturally accommodated by robust smooth pulses \cite{Pryadko2008,Barnes2015,Zeng2019,Throckmorton2019}. These smooth pulses have an analytical form, but with free parameters that must be chosen to produce the desired unitary while satisfying robustness constraints, and this usually requires a numerical search in parameter space.

In this paper, based on the approach of Ref.~\cite{Barnes2015}, we derive a completely analytical family of robust smooth pulses which eliminates the requirement of numerical parameter fitting. We also cast the problem of finding a robust smooth pulse which implements a particular unitary into a set of coupled ordinary differential equations (ODEs), which can be solved by using standard numerical solvers. We provide explicit examples of robust pulse shapes along with their filter functions.

Although our focus here will be on a two-level system, the physical context is not necessarily limited to one-qubit problems. Indeed, these solutions can be used to implement robust gates in SU(2) $\subset$ SU(4) or SU(2)$\times$SU(2) $\subset$ SU(4) subgroups, targeting local rotations or non-local controlled-phase gates in a silicon double quantum dot setup \cite{Gungordu2018a} or in superconducting qubits with fixed coupling \cite{Chow2011}.

The structure of this paper is as follows.  In Sec.~\ref{sec:Background}, we present a brief summary of the analytical formalism of Barnes et al.~\cite{Barnes2015} on which this work is built.  In Sec.~\ref{sec:analytical}, we show how to choose symmetric auxilliary functions and their parameters without resorting to a numerical search, and we present the resulting pulse shapes and filter functions.  In Sec.~\ref{sec:ODEs}, we show how to efficiently generate robust pulse shapes by introducing auxilliary ODEs and incorporating the desired rotation angles and robustness constraints as local boundary conditions rather than nonlocal integral relations.
We then conclude in Sec.~\ref{sec:conclusions}.

\section{Background}\label{sec:Background}

We first review robust smooth pulses for a two-level system, adapted from Ref.~\cite{Barnes2015} to our use cases.
We consider the Hamiltonian
\begin{align}
\tilde H(t) =  \Omega_0(t) \sigma_z + \tilde \beta \sigma_x,
\label{eq:H}
\end{align}
where $\Omega_0(t)$ represents the driving field, and $\tilde \beta = \beta + \delta \beta$ is the energy splitting with non-Markovian fluctuations $\delta \beta$.

This Hamiltonian appears in various systems including solid state spin qubits, and thus our results have a wide applicability.
For the sake of having specific numbers and constraints for the Hamiltonian and noise levels, however, we remark that our main interest is spin qubits in a double quantum dot setup in the (1,1) charge configuration, for which the exchange coupling $J$ between the electrons remains fixed and one of the electrons is driven by a microwave source whose amplitude is given by $\Omega_\text{mw}$.  This mapping requires that the Zeeman splittings of the electrons are sufficiently different from each other such that it is possible to address one of the electrons without affecting the other, as is the case in the experiments of Refs. \cite{Veldhorst2014a,Zajac2017a,Watson2017}. Under these conditions, the noisy Hamiltonian for the spin pair can be expressed as
\begin{align}
\tilde H(t) = \frac{\Omega_\text{mw}(t)}{2} IX + \frac{\tilde J}{4} ZZ,
\end{align}
where $\Omega_\text{mw}(t)$ is proportional to the amplitude of the microwave drive, $\tilde J = J + \delta J$ and $\delta J$ denotes the noise in the exchange coupling due to electrostatic fluctuations \cite{Gungordu2018a}. This Hamiltonian is in an $\mathfrak{su}(2)$ subalgebra of $\mathfrak{su}(4)$ generated by $\{IX,ZZ,ZY\}$, and thus the results we obtain below for the Hamiltonian in Eq.~\eqref{eq:H} can be applicable to these devices.

We now summarize the formalism from Ref.~\cite{Barnes2015} for finding robust pulse shapes to fix quasistatic stochastic errors in $\beta$ while targeting a specific rotation at the final time $t_f$.
 In the absence of noise, the time evolution operator $U(t_f;0)$ at $t=t_f$ can be parametrized in terms of an auxiliary function $\Phi(\chi)$ (where $\chi = \chi(t)$ is a reparametrization of time), expressed via the $ZXZ$ Euler angle decomposition in the following way \cite{Barnes2015}:
\begin{align}
U(t_f) = Z_{\xi_+(\chi_f) - \xi_-(\chi_f)} X_{2\chi_f} Z_{-[\xi_+(\chi_f) + \xi_-(\chi_f)]}
\label{eq:euler}
\end{align}
where $\chi_f = \chi(t_f)$, and $X_\gamma$ ($Z_\gamma$) denotes a rotation around the $x-$ ($z-$) axis of the Bloch sphere by angle $\gamma$. $\xi_{\pm}$, which determine the Euler angles, are related to the parametrizing function $\Phi(\chi)$ through
\begin{align}
\label{eq:xi}
\xi_\pm(\chi_f) = &\Phi(\chi_f) \mp \\
&\text{sgn}(\Phi'(\chi_f)) \frac{1}{2}\text{arcsec}\left( \sqrt{ 1 + [\Phi'(\chi_f) \sin(2\chi_f)]^2 } \right). \nonumber
\end{align}
The control field $\Omega_0(t)$ is related to $\Phi(\chi)$ through
\begin{align}
\label{eq:Omega}
\Omega_0(t) = \Omega(\chi) =& -\beta \sin(2\chi) \times \\
&\frac{\Phi''(\chi)  + 4\Phi'(\chi)\cot(2\chi) +[\Phi'(\chi)]^3 \sin(4\chi)}{ 2 \sqrt{1+\left[\Phi'(\chi)\sin(2\chi)\right]^2 }^3 }, \nonumber 
\end{align}
and $\chi$ is a reparametrization of time, determined by $\Phi(\chi)$ \cite{Barnes2015} as follows:
\begin{align}
\beta t = \hbar \int_{0}^{\chi} d\bar\chi \sqrt{1+\left[\Phi'(\bar\chi)\sin(2\bar\chi)\right]^2 }.
\label{eq:tf}
\end{align}
We note that the initial condition $U(0)=\openone$ implies $\Phi(0)=\Phi'(0)=0$.

The main point of the above is that the problem of finding a pulse shape, which will result in a target rotation $U(t_f) = U_\text{target}$ at time $t=t_f$ in the absence of noise can be reduced to choosing a function $\Phi(\chi)$ which only needs to obey certain \emph{local} boundary conditions (rather than \emph{nonlocal} integral conditions as in Refs.~\cite{Barnes2012,Barnes2013,Zeng2019}), and the choice of $\Phi(\chi)$ in turn determines the pulse shape $\Omega_0(t)$ that needs to be applied during this time interval to make this happen.

Now, to ensure that this time-evolution is also robust against quasistatic stochastic noise in $\beta$, which is our goal in this work as well, $\Phi(\chi)$ should further satisfy the following additional conditions \cite{Barnes2015}:
\begin{align}
\epsilon_\chi(\chi_f) \equiv& \sin^2(2\chi_f) e^{2 i \Phi(\chi_f) }   + \nonumber\\
& 4\tan(2\chi_f) \int_{0}^{\chi_f} d\chi \sin^2(2\chi) e^{2 i \Phi(\chi) } = 0 \label{eq:robustness1} \\
\epsilon_\xi(\chi_f) \equiv& \int_{0}^{\chi_f} d\chi \sin^2(2\chi) \Phi'(\chi) = 0.
\label{eq:robustness2}
\end{align}
These relations follow from the series expansion of noisy time evolution operator in powers of noise terms. The real and imaginary parts of the left hand side of Eq.~\eqref{eq:robustness1} are proportional to the coefficients for the leading order noise terms for $\delta_\beta \chi(t_f)$ and $\delta_\beta \dot\chi(t_f)$, respectively, and the left hand side of Eq.~\eqref{eq:robustness2} gives the coefficient for $\delta_\beta\xi(t_f)$ \cite{Barnes2015}, and they need to vanish at the final time $\chi = \chi_f$ such that the gate is robust.

The second of these robustness conditions is automatically satisfied for any driving pulse that is antisymmetric (odd) in time: When $\Phi(\chi)$ is an even function, Eq.~\eqref{eq:robustness2} is odd in $\chi_f$ (which itself is an odd function of $t$ which follows from Eq.~\eqref{eq:tf}), so for any choice of $\Phi(\chi)$ on the interval $\left[0,\chi_f\right]$, or correspondingly, $\Omega_0(t)$ on $\left[0,t_f\right]$, one can construct a rotation robust against $\delta_\beta\xi(t_f)$ by extending the evolution to the symmetric interval $\left[-t_f,t_f\right]$ with $\Omega_0(t)$ at negative times defined by enforcing antisymmetry $\Omega_0(-t) = -\Omega_0(t)$.  One can find the resulting overall time evolution $U(t_f;-t_f)$ as follows. From Eq.~\eqref{eq:H} and the Schr\"odinger equation $i \dot U(t) = H(t) U(t)$, we see that a similarity transformation by $\sigma_z$ combined with a time-inversion of a time-evolution with $H(t)$ from $t=0$ to $t_f$ is equivalent to a time-evolution by $H'(t) = -\Omega_0(t) + \beta$ from $t=0$ to $t=t_f$. Thus the time evolution for the first half with ``inverted" pulse can be written in terms of the time evolution for the second half $U(t_f;0)$ as $\sigma_z U_f(t_f;0)^\dagger \sigma_z$, which leads to
\begin{align}
U(t_f;-t_f) = U(t_f;0) [\sigma_z U(t_f;0)^\dagger \sigma_z],
\end{align}
and finally, by using Eqns.~\eqref{eq:euler} and ~\eqref{eq:xi}, we obtain
\begin{align}
U(t_f;-t_f) = Z_{\xi_+(\chi_f) - \xi_-(\chi_f)} X_{4\chi_f} Z_{-[\xi_+(\chi_f) - \xi_-(\chi_f)]},
\end{align}
which is a rotation on the Bloch sphere given by an angle $\theta = 4\chi_f$ around an axis $\cos(\phi) \hat{x} + \sin(\phi)\hat{y}$ where \cite{Barnes2015}
\begin{align}
\cos\phi=\frac{1}{ \sqrt{1+\left[\Phi'(\chi_f)\sin(2\chi_f)\right]^2 } }.
\label{eq:axis}
\end{align}
When the bandwidth on the control field $\Omega_0(t)$ is limited such that it cannot be turned on or off quickly (when compared to the timescale $t_f$), one can furthermore require that $\Omega_0(t_f)$ also vanishes, which can be viewed as a constraint on $\Phi''(\chi_f)$ \cite{Barnes2015} via Eq.~\eqref{eq:Omega}.

The first of the robustness conditions, the complex-valued Eq.~\eqref{eq:robustness1}, however, cannot be as trivially satisfied.  When targeting an arbitrary rotation, it is possible to find solutions by starting with an ansatz for the auxilliary function $\Phi(\chi, \boldsymbol a)$ with sufficient degrees of freedom encapsulated as $\boldsymbol a$, and use a numerical search to find $a_i$ which would satisfy the robustness conditions while at the same time producing the desired rotation \cite{Barnes2015}. For the special case of $\chi_f = n\pi/4$, analytical solutions were given in \cite{Barnes2015}. In the next section, we show how to satisfy Eq.~\eqref{eq:robustness1} analytically for an \emph{arbitrary} unitary.

\section{Analytical solutions to the robustness conditions}\label{sec:analytical}
Within this section, we assume that Eq.~\eqref{eq:robustness2} will be satisfied by doubling the interval to $\left[-\chi_f,\chi_f\right]$ and using symmetry as discussed above.  Thus, we only need to focus on satisfying Eq.~\eqref{eq:robustness1}.  This will ensure that the strength of the leading order noise term in the time-evolution operator vanishes at the final time, which is parametrized as $\chi_f$.

Now we will replace the auxiliary function $\Phi(\chi)$ with two new auxiliary functions (as one can already surmise, the two new functions will not be independent of each other), $\mathcal R(\chi)$ and $\alpha(\chi)$, defined via
\begin{equation}\label{eq:Ra}
\mathcal R(\chi) e^{i \left[2\Phi(\chi)-\alpha(\chi)\right]} \equiv \int_0^{\chi} d\bar{\chi} \sin^2(2\bar{\chi}) e^{2 i \Phi(\bar{\chi}) }.
\end{equation}
Differentiating both sides of Eq.~\eqref{eq:Ra} with respect to $\chi$ one can obtain
\begin{equation}\label{eq:alpha}
e^{i \alpha(\chi) }\sin^2(2\chi) = \mathcal R'(\chi) + i \mathcal R(\chi) [2 \Phi'(\chi)-\alpha'(\chi)],
\end{equation}
and so, for a given $\alpha(\chi)$, solving the real and imaginary parts of Eq.~\eqref{eq:alpha} gives the relations
\begin{align}
\label{eq:R}
\mathcal R(\chi) &= \int_0^\chi d\bar\chi \cos(\alpha(\bar\chi)) \sin^2(2\bar\chi),\\
\Phi'(\chi) &=  \frac{1}{2}\left[ \alpha'(\chi) + \frac{\sin(\alpha(\chi)) \sin^2(2\chi)}{ \mathcal R(\chi) } \right] \nonumber.
\end{align}
The first relation indicates that $\mathcal R(\chi)$ and $\alpha(\chi)$ are not independent; either one can be used to parametrize the time evolution operator.
The second relation tells us how to translate a pulse specified in the $\mathcal R(\chi)$/$\alpha(\chi)$ parametrization back to the original parametrization in Ref.~\cite{Barnes2015} in terms of $\Phi(\chi)$.

At this point, we have reparametrized the solution of the Schr\"odinger equation in terms of a function $\alpha(\chi)$ (or  equivalently $\mathcal R(\chi)$) instead of $\Phi(\chi)$, and showed how it would be related to $\Phi(\chi)$ of the original parameterization. The advantage of this reparametrization is that the robustness conditions simplify into \emph{local} boundary conditions for $\alpha(\chi)$ and $\mathcal R(\chi)$ as opposed to \emph{nonlocal} robustness conditions on $\Phi(\chi)$, as we will show shortly.

Furthermore, the problem is analytically solvable when $\alpha(\chi)$ is chosen in such a way that $\cos\alpha(\chi) \sin^2(2\chi)$ is integrable. A better alternative, however, is to consider $\mathcal R(\chi)$ as the independent variable, which in turn defines $\alpha(\chi)$ through its derivative.

Before going into the robustness conditions, though, we note that the condition $\Phi(0)=0$ (from the initial condition $U(0)=\openone$) simply corresponds to a vanishing integration constant in Eq.~\eqref{eq:R}. However, the condition $\Phi'(0) = 0$ (also from $U(0)=\openone$) requires special care: since the denominator vanishes in the limit $\chi \to 0$ (and possibly at other points, depending on the choice for $\alpha(\chi)$), we impose $\sin(\alpha(\chi)) =0$ at these points to avoid any singularities. This is a stronger condition than requiring that the strength of the control pulse $|\Omega(\chi)|$ remains finite, but leads to a simpler set of constraints.

Finally, we can relate the robustness condition, Eq.~\eqref{eq:robustness1}, to boundary conditions on $\alpha(\chi)$ and $\mathcal R(\chi)$ by noting that
\begin{align}\label{eq:Ra-def}
 \left(\mathcal R(\chi) e^{i \left[2\Phi(\chi)-\alpha(\chi)\right]}\right)' + 4 \tan(2\chi)\left(\mathcal R(\chi) e^{i \left[2\Phi(\chi)-\alpha(\chi)\right]} \right)
 = \epsilon_\chi(\chi).
\end{align}
Plugging Eq.~\eqref{eq:R} into the lhs and separating real and imaginary parts yields
\begin{align}
\text{Re}\left(e^{i\alpha(\chi)-2i\Phi(\chi)} \epsilon_\chi(\chi)\right) =& 4 \mathcal R(\chi) \tan(2\chi) + \cos(\alpha(\chi)) \sin^2(2\chi), \nonumber\\
\text{Im}\left(e^{i\alpha(\chi)-2i\Phi(\chi)} \epsilon_\chi(\chi)\right) =& \sin(\alpha(\chi)) \sin^2(2\chi).
\end{align}
So, for a generic value $\chi_f$ (recalling that it is one of the Euler angles of the final rotation and hence should not be restricted), the robustness condition $\epsilon_\chi(\chi_f) = 0$ reduces these relations to
\begin{align}
\alpha(\chi_f) = n \pi, \qquad \mathcal R(\chi_f) =  \frac{(-1)^{n}}{8}\sin(4\chi_f),
\label{eq:bcf-alpha}
\end{align}
where $n$ is any integer.

At this point, besides these boundary conditions, we remark that if we treat $\mathcal R(\chi)$ as the fundamental parameterizing function of the problem, it still cannot be chosen arbitrarily during the intermediate times $\chi \in (0,\chi_f)$, because Eq.~\eqref{eq:R} implies that its derivative must be a function within $[-1,1]$ at all times. Also, ensuring that $\Phi'(\chi)$ remains finite requires special care; a straightforward way to achieve this is to require that
\begin{align}
|\mathcal R(\chi)| &> 0,  \quad \forall \chi \in (0,\chi_f], \nonumber \\
|\mathcal R'(\chi)| &< 1, \quad \forall \chi \in (0,\chi_f).
\label{eq:finite}
\end{align}
Furthermore, the robustness condition Eq.~\eqref{eq:bcf-alpha} for $\alpha(\chi)$ translates into a boundary condition on  $\mathcal R'(\chi)/\sin^2(2\chi)$.

The relation between the phase, $\alpha(\chi)$, and the amplitude $\mathcal R(\chi)$ can be slightly simplified by introducing yet one more parametrization of time $u(\chi) \equiv 4\chi - \sin(4\chi)$ and define $\beta(u(\chi)) \equiv \alpha(\chi)$, such that $\mathcal R(u) = \int_0^u du \cos\beta(u)$. Using $u \equiv u(\chi)$ instead of $\chi$ as the ``independent" parameter is not essential and working with $\mathcal R(\chi)$ is equally possible and will be used in Section \ref{sec:ODEs} when finding solutions numerically, but we find it convenient when looking for analytical solutions.

Under this reparameterization, $\Phi'(\chi)$ simplifies to
\begin{align}
\Phi'(\chi) = \frac{u'(\chi)}{2}\left[ -\frac{\mathcal R''(u)}{\sqrt{1-\mathcal R'(u)^2}} + \frac{\sqrt{1-\mathcal R'(u)^2}}{\mathcal R(u)} \right]
\label{eq:Phi}
\end{align}
and the problem of finding a robust unitary reduces to picking a function $\mathcal R(u)$ which satisfies the following ordinary relations:
\begin{align}
\mathcal R'(u=u_f) = \pm 1, \quad \mathcal R(u=u_f) = \mp \sin(4\chi_f)
\label{eq:Fbc2}
\end{align}
for robustness, and 
\begin{align}
&\mathcal R'(u=0) = \pm 1, \qquad \mathcal R'(u) \in [-1,1]
\label{eq:Fbc1} \\
&\mathcal R''(u=0,u_f)=0, \nonumber
\end{align}
by construction, and
finally Eq.~\eqref{eq:axis}, or equivalently
\begin{align}
\tan^2\phi = \lim_{\chi \to \chi_f} - 16  \sin^6(2\chi) \text{sgn}[\mathcal R'(u)] \mathcal R'''(u) ,
\label{eq:axisF}
\end{align}
for targeting the unitary
\begin{align}
U(\theta,\phi) = e^{-i \frac{\theta}{2} (\cos\phi \sigma_x + \sin\phi \sigma_y) }
\label{eq:U}
\end{align}
at the final time, where $\theta = 4\chi_f$.
The first set of boundary conditions above, for $\mathcal R'(u=0)$ and $\mathcal R''(u=0,u_f)$, are required to ensure $\Phi'(\chi \to 0)$ (hence the driving field) remains finite, since the denominator $\mathcal R(u)$ in Eq.~\eqref{eq:Phi} vanishes, and the remaining two follow from the robustness requirement.

Finally, $\Omega(\chi)$ is an odd function when $\mathcal R(u)$ is also odd, which ensures that the second robustness condition Eq.~\ref{eq:robustness2} is satisfied when pulsing in a symmetric time interval.

In summary, for a given $\theta$ and $\phi$, any choice of function $\mathcal R(\chi)$ whose derivative is bounded in the interval $[-1,1]$ and which obeys the boundary conditions given in Eqns.~(\ref{eq:Fbc2}--\ref{eq:axisF}), and Eq.~\eqref{eq:finite} will yield a pulse shape $\Omega(\chi)$ via Eqns.~\eqref{eq:Omega} and \eqref{eq:Phi}, which results in the robust quantum gate $U(\theta,\phi)$ (cf. Eq.~\eqref{eq:U}) implemented using the Hamiltonian $H(t)$ (cf. Eq.~\eqref{eq:H}). Different choices for $\mathcal R(\chi)$ will yield different pulse shapes but the resulting robust gate will be the same.

\subsection{Examples}
\begin{figure}[h]
(a) \includegraphics[width=0.45\textwidth]{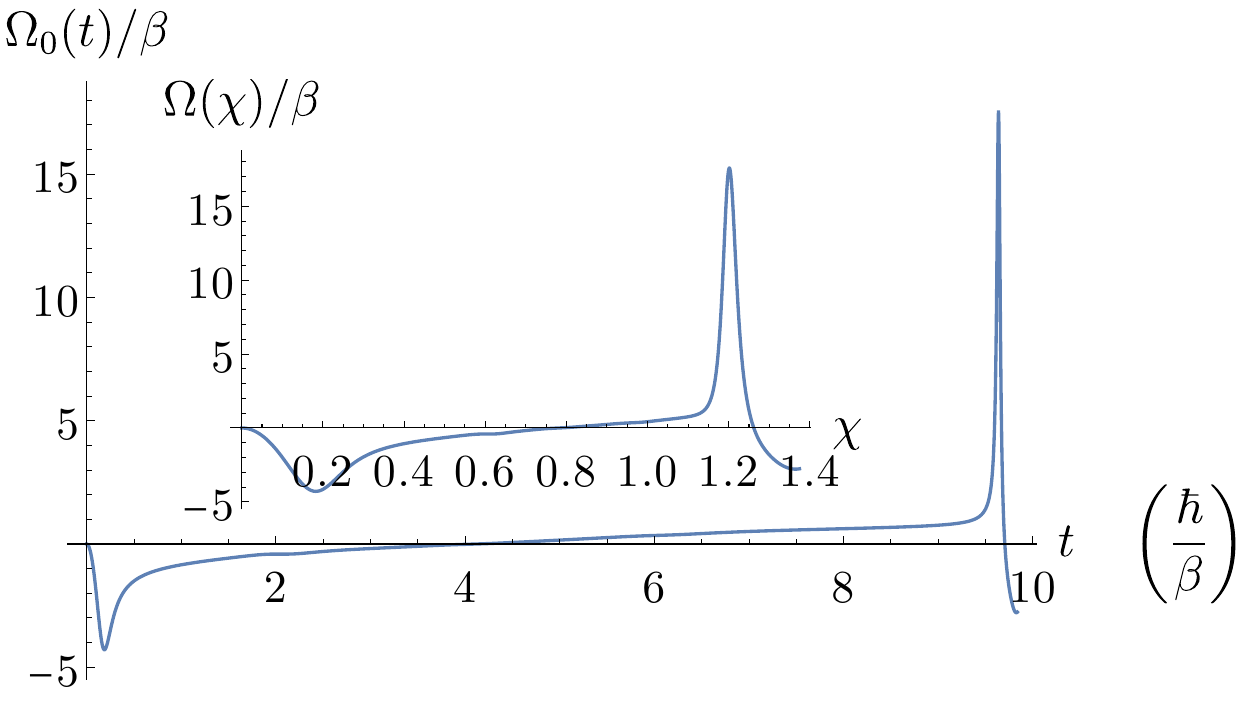}\\
(b) \includegraphics[width=0.40\textwidth]{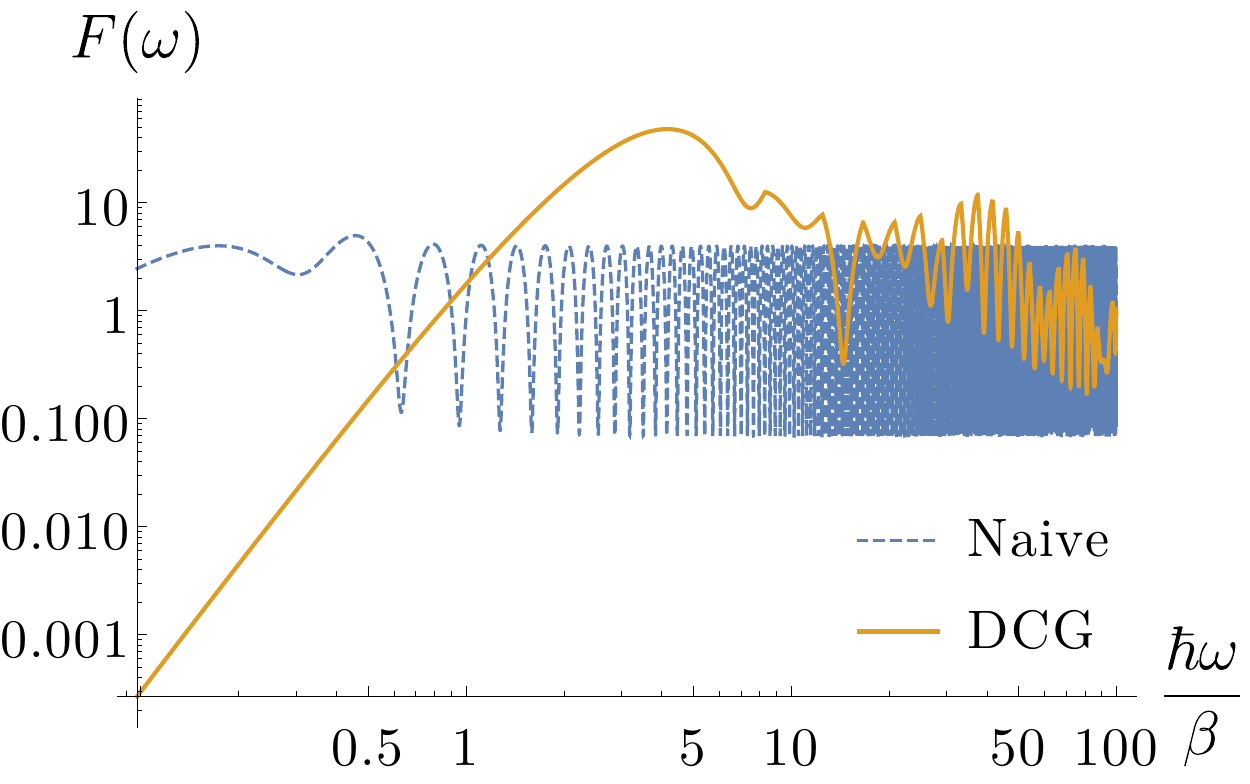}\\
\caption{(Color online) (a) Pulse shape $\Omega_0(t)$ (inset $\Omega(\chi)$) in units of $\beta$ which implements a robust $\theta = 2\pi-\pi/2$ rotation around the axis $\boldsymbol n = (\cos\phi,\sin\phi,0)$ with $\phi = \pi/9$.
(b) Comparison of the leading order filter functions for the robust gate against a naive implementation using $U_\text{naive}(t)$.
}
\label{fig:ex1}
\end{figure}

\begin{figure}[h]
(a) \includegraphics[width=0.45\textwidth]{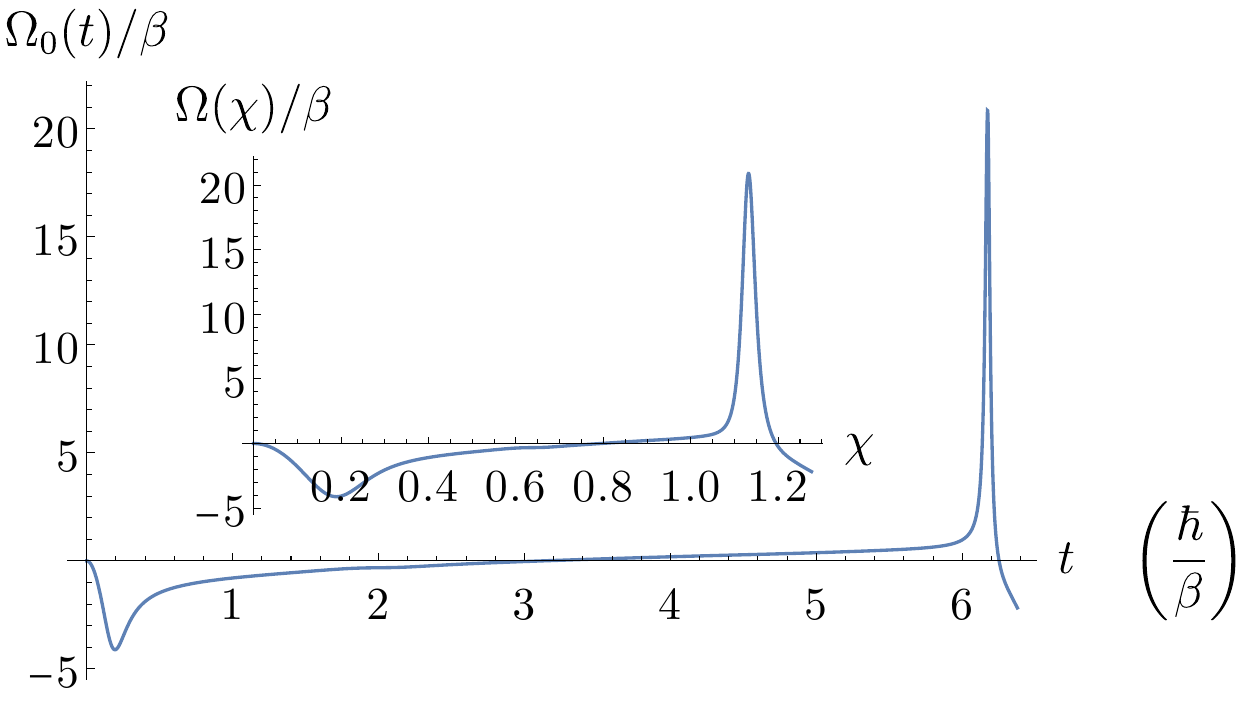}\\
(b) \includegraphics[width=0.4\textwidth]{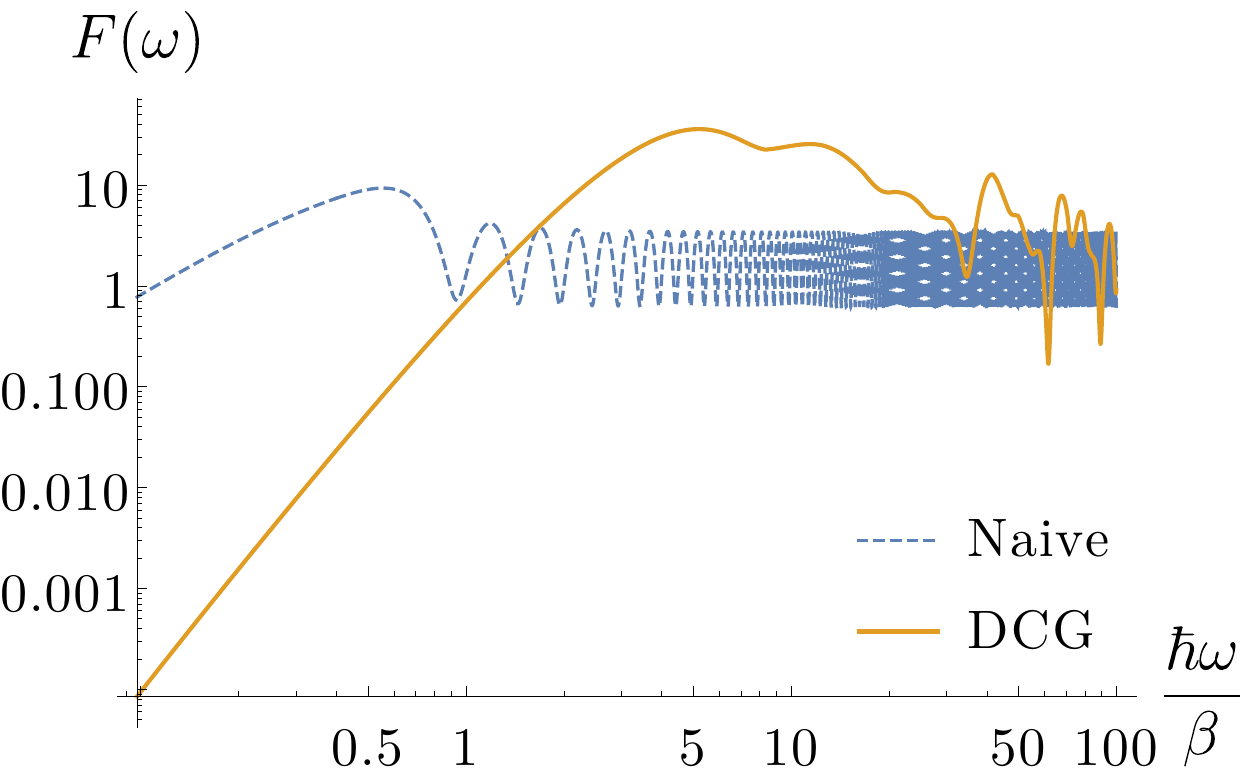}\\
\caption{(Color online) (a)  Pulse shape $\Omega_0(t)$ (inset $\Omega(\chi)$) in units of $\beta$ which implements a robust $\theta =  2\pi-3\pi/8$ rotation around the axis $\boldsymbol n = (\cos\phi,\sin\phi,0)$ with $\phi = \pi/4$.
(b) Comparison of the leading order filter functions for the robust gate against a naive implementation using $U_\text{naive}(t)$.
}
\label{fig:ex2}
\end{figure}

As an example, consider the following even function:
\begin{align}
\mathcal R'(u) = a_0 + a_1 \cos\left(\frac{2\pi u}{u_f}\right) + a_2 \cos\left(\frac{4\pi u}{u_f}\right).
\label{eq:ex-ansatz}
\end{align}
which is trivially integrable, and $\mathcal R(u_f)$ is simply given by $a_0 u_f$ since oscillatory functions integrate to zero at the final time. And by letting $a_0 = -\sin(4 \chi_f)/u_f$ and $a_2 = 1 - a_0 - a_1$, we meet all the robustness conditions in Eq.~\eqref{eq:Fbc2}.

We can target, say, a $\theta = 4\chi_f = 2\pi - \pi/2$ rotation around the axis given by $\phi = \pi/9$ using Eq.~\eqref{eq:axisF}, which corresponds to the choice $a_1 \approx 0.3244$. From Eq.~\eqref{eq:tf}, we find the total gate time is $t_f \approx 9.84 \hbar/\beta$. The resulting pulse shape is shown in Fig.~\ref{fig:ex1}. Similarly, for $\theta = 2\pi-3\pi/8$, $\phi=\pi/4$, we find $a_1 \approx 0.4767$ and obtain $t_f \approx 6.38 \hbar/\beta$ (Fig.~\ref{fig:ex2}).

\begin{figure}[h]
\includegraphics[width=0.5\textwidth]{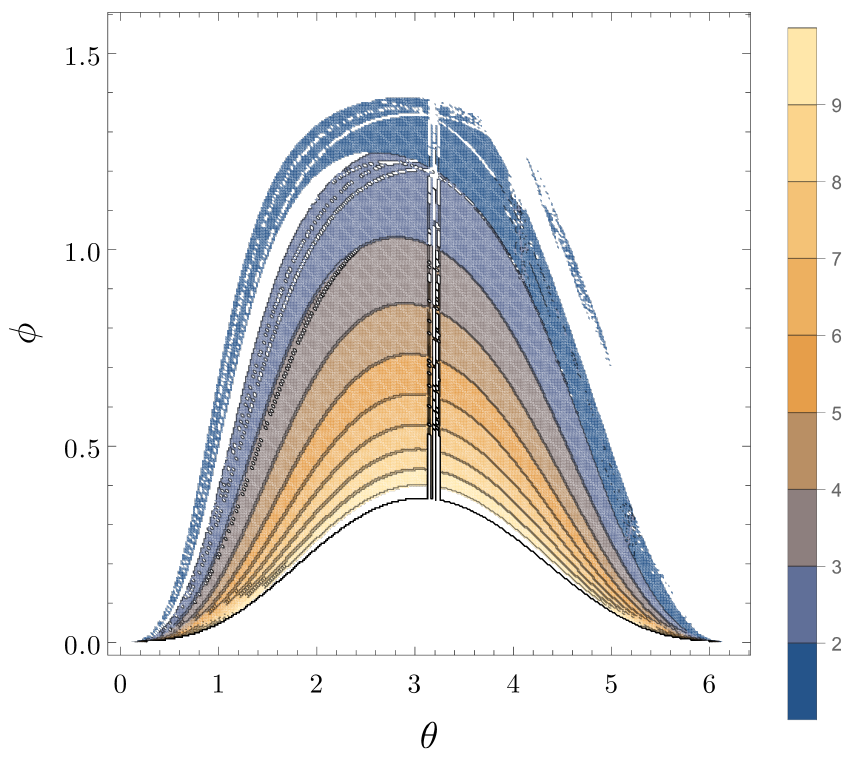}\\
\caption{(Color online) Accessible unitaries $U(\theta + 2\pi n,\phi)$ when using the ansatz from Eq.~\eqref{eq:ex-ansatz}. The minimum number of additional windings to target the unitary, $n$, is color coded, up to $n=10$. White regions either require $n>10$ or cannot be implement with this ansatz, either because it leads to solutions for which $\mathcal R(u)$ vanishes for some $u>0$ which implies a divergent $\Phi'(\chi)$, or because $\mathcal R'(u)$ exceeds the interval $[-1,1]$ for some $u \in (0, u_f)$.}
\label{fig:ex-ansatz-windings}
\end{figure}

When using this ansatz, targeting other unitaries may require additional $2\pi$ windings in $\theta$. The minimum number of additional windings required for targeting an arbitrary unitary $U(\theta,\phi)$ is shown in Fig.~\ref{fig:ex-ansatz-windings}. Overall, these pulses require a bandwidth of $\sim 100 \beta/\hbar$ when targeting fidelities above 99.99\%.
For example, a typically accessible bandwidth of 40MHz \cite{Yang2019} limits $\beta/h$ to $0.4$MHz, which implies an exchange coupling of $J/h =  1.6\text{MHz}$, and a maximum microwave amplitude of $\text{max}[\Omega_\text{mw}(t)/h] \approx 16$MHz for the pulse given in Fig.~\ref{fig:ex2}, which is attainable in the experiments \cite*{Zajac2017a,Russ2017a}. The resulting gate time $t_f \approx 2.54\mu$s is less than $T_2^* \sim 10\mu$s \cite{Veldhorst2014a}, and orders of magnitude less than $T_2$, which is the relevant time scale for dynamical error correction.

It is possible to impose additional constraints, such as $\Omega_0(t_f)=0$ to soften the tail, using this form of ansatz, although this requires adding higher harmonics with free coefficients of the form $a_m \cos(2 m \pi u /u_f)$.

We conclude this section by remarking that finding an ansatz function $\mathcal R(\chi)$ with sufficient number of tunable parameters $a_i$ and which meets the conditions Eqns.~(\ref{eq:finite}--\ref{eq:Fbc2}) is not easy to come by.
For example, the ansatz Eq.~\eqref{eq:ex-ansatz} requires careful tuning of $a_i$ to ensure that $\mathcal R'(u) \in (-1,1)$ for all $u \in (0,u_f)$ is satisfied and that the denominators in Eq.~\eqref{eq:Phi} remain nonzero; the white regions in Fig.~\ref{fig:ex-ansatz-windings} contain the parameter regions which fail this check. These continuous conditions can in principle be satisfied by construction with better choice of ansatz. For example, the simple choice for the even function $\mathcal R'(u) = \cos^2\left(\sum_{i=0}^N a_i u^{2i}\right)$ with $N \geq 3$ does satisfy these continuous constraints from the outset, but unfortunately, it is not possible to analytically integrate this function to find $\mathcal R(u)$ for $N>2$. A numerical approach is still viable, however, which we will demonstrate in Section \ref{sec:ODEs}.

\subsection{Filter function}
The smooth pulse is designed to cancel quasistatic noise, i.e., noise that is constant during the gate duration $t \in [-t_f, t_f]$. In practice, the noise strength may also drift during the pulse. For instance, in the context of the double quantum dot setup in \cite{Gungordu2018a}, $\Omega_0(t)$ corresponds to the microwave driving amplitude and $\beta$ error corresponds to exchange error induced by charge noise, which typically has a $1/f$ power spectral density (PSD) in the relevant region of the noise spectrum. When the noise is sufficiently weak such that the error Hamiltonian $H_\epsilon(t)$ satisfies $||\int_{t_0}^{t_f} dt H_\epsilon(t)|| \ll 1$, the average susceptibility of a quantum gate to time-dependent noise can be characterized in a perturbative manner. In this approach, the leading order error in noise-averaged fidelity is given by
\begin{align}
\langle \mathcal F \rangle \approx 1- \frac{1}{\hbar^2} \sum_{i,j=1}^3 \int_{-\infty}^{\infty} \frac{d \omega}{2\pi} S_{ij}(\omega)\frac{F_{ij}(\omega)}{\omega^2}.
\end{align}
where $S(\omega)$ and $F(\omega)$ respectively characterize the noise and control, and are related to the error and control Hamiltonians ($H_\epsilon(t) = \delta \beta(t) \sigma_x$ and $H_c = \Omega_0(t) \sigma_z + \beta \sigma_x$ in our case) as follows. The filter function, $F(\omega)$ is given by
\begin{align}
F(\omega) = [R(\omega) R^\dagger(\omega)]^T
\end{align}
where $R_{ik}(\omega) \equiv -i \omega \int_{t_0}^{t_f} dt  R_{ik}(t)e^{i \omega t}$ and $R(t) = \text{Ad}(U(t;t_0)) = \text{tr}(\sigma_i U(t;t_0) \sigma_j U^\dagger(t;t_0))/2$ is the adjoint representation of the time-evolution operator \cite{Green2013,Gungordu2018a}.
$S_{ij}(\omega)$ is the power spectral density given by Fourier transforming the correlation between the coefficients of $\sigma_i$ and $\sigma_j$ terms in the noise Hamiltonian $H_\epsilon(t)$. In our particular case, only $S_{xx}(\omega)$ is non-zero and is given by the Fourier transform of the autocorrelation function $C_\beta(t)=\langle \delta\beta(t) \delta\beta(0) \rangle$.

We have numerically evaluated the filter functions corresponding to the gates obtained by the control pulses given in Figs.~\ref{fig:ex1} and \ref{fig:ex2} in a symmetric time interval from $-t_f$ to $t_f$. We compare their filter function to that of a naive pulse
\begin{align}
U_\text{naive}(t) = \exp\left[-i \theta\frac{t+t_f}{2 t_f} (\cos\phi\sigma_x + \sin\phi \sigma_y)\right]
\end{align}
which also implements the same unitary in the same amount of time. The results are shown in Figs.~\ref{fig:ex1} and \ref{fig:ex2}. The robust gates suppress the low frequency noise much better than the corresponding naive gates, although they are more susceptible to noise at frequencies on the order of inverse gate time, $\omega \sim 1/t_f$. Thus the dynamically corrected gates (DCGs) tend to lead to higher fidelities when the noise power is concentrated at frequencies lower than $\omega \sim 1/t_f$, which is the case for these devices \cite*{Yoneda2017a,Chan2018a,Gungordu2019a}.

\section{Robust pulse shapes as solutions of coupled ODE systems}\label{sec:ODEs}
In this section, we show that the problem of finding robust pulse shapes can be converted into a set of coupled ODEs. This allows finding more general solutions, which are even not restricted to antisymmetric pulse shapes in principle, by using standard ODE solvers in a straightforward manner. This method still avoids any search over parameters, and yields solutions very quickly.

We first make a change of variables to ensure that denominator in Eq.~(\ref{eq:Phi}) never vanishes for $\chi > 0$. A straightforward way of achieving this would be to ensure that the integrand of the denominator is always positive (or negative), which can be achieved by defining yet another function which is bounded, $\gamma(\chi)$, such that
\begin{align}
A \tanh(\gamma(\chi)) \equiv \alpha(\chi), \qquad \pi/2 \geq A > 0.
\end{align}
In terms of $\gamma(\chi)$, the robustness conditions then become
\begin{align}
\gamma(0) = 0, \quad \gamma(\chi_f) = \tanh^{-1}(\alpha(\chi_f)/A), \quad \gamma'(0) = 0
\label{eq:}
\end{align}
and we can solve for the second condition in Eq.~(\ref{eq:bcf-alpha}) by considering a differential equation
\begin{align}
G'(\chi) = \cos(A \tanh(\gamma(\chi))) \sin^2(2\chi),
\label{eq:G}
\end{align}
subject to boundary condition
\begin{align}
G(\chi_f) = -\frac{1}{8}\cos(\alpha(\chi_f))\sin(4\chi_f).
\end{align}
This function must also satisfy
\begin{align}
G(0) = 0
\end{align}
since $\mathcal R(0)=0$.

The rotation axis defined by the angle $\phi$ can be imposed via a boundary condition on $\alpha(\chi_f)$, using Eqns.~\eqref{eq:axis}, \eqref{eq:R} and \eqref{eq:bcf-alpha}, which gives:
\begin{align}
\alpha'(\chi_f) = 2\tan(\phi)/\sin(2\chi_f).
\end{align}

We can also impose the condition that the pulse $\Omega(\chi)$ should vanish at the end ($\Omega(\chi_f) = 0$) by imposing a boundary condition on $\alpha''(\chi_f)$, using Eq.~(\ref{eq:Omega}):
\begin{align}
\Phi''(\chi_f) =&  - 4\Phi'(\chi_f)\cot(2\chi_f) -[\Phi'(\chi_f)]^3 \sin(4\chi_f).
\end{align}
Since $\Phi'(\chi_f) = \alpha'(\chi_f)/2$ and $\Phi''(\chi_f) = [\alpha''(\chi_f) -4 \alpha'(\chi_f) \tan(2\chi_f)]/2$, this can be seen as the defining condition on $\alpha''(\chi_f)$.

These two boundary conditions on the first and second derivatives of $\alpha(\chi_f)$ can readily written as corresponding boundary conditions on the derivatives of $\gamma(\chi)$, as
\begin{align}
\gamma'(\chi_f) &= \frac{\alpha'(\chi_f)}{A \left(1-\frac{\alpha^2(\chi_f)}{A^2} \right)} \nonumber\\
\gamma''(\chi_f) &= 2 \frac{\alpha(\chi_f) \alpha'(\chi_f)^2}{ A^3 \left(1-\frac{\alpha^2(\chi_f)}{A^2} \right)^2 } + \frac{\alpha''(\chi_f)}{A \left(1-\frac{\alpha^2(\chi_f)}{A^2} \right)}
\end{align}

\begin{figure}
(a) \includegraphics[width=0.45\textwidth]{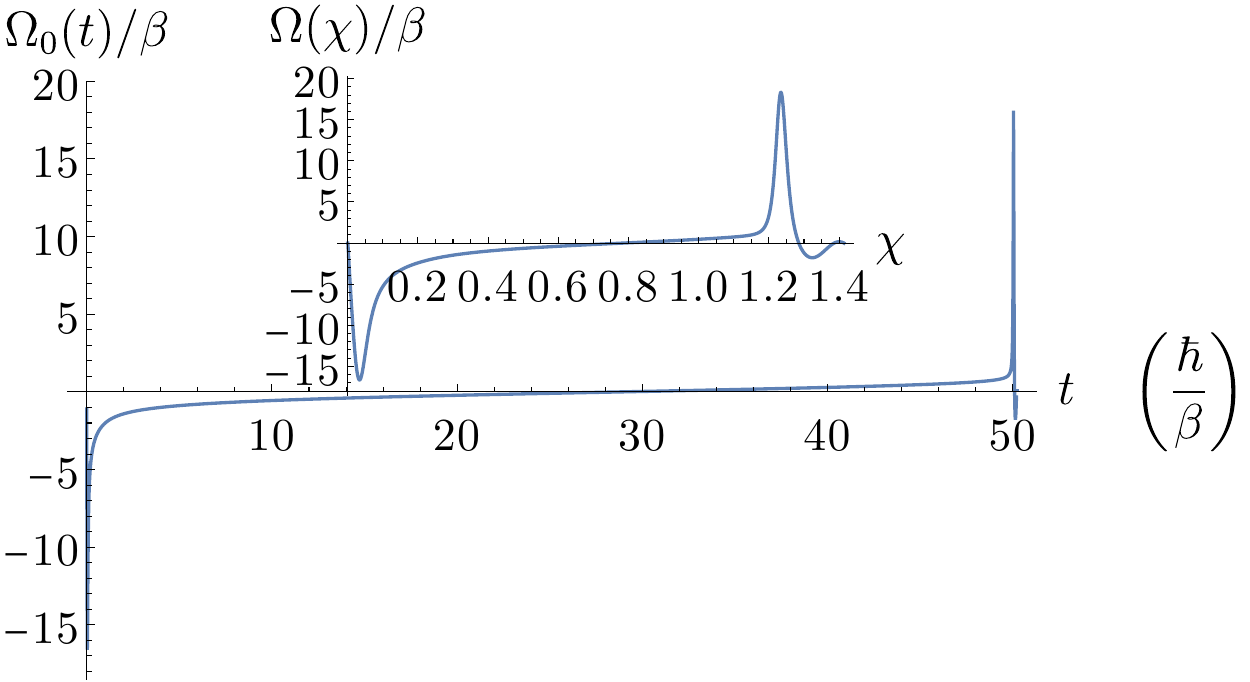}\\
(b) \includegraphics[width=0.4\textwidth]{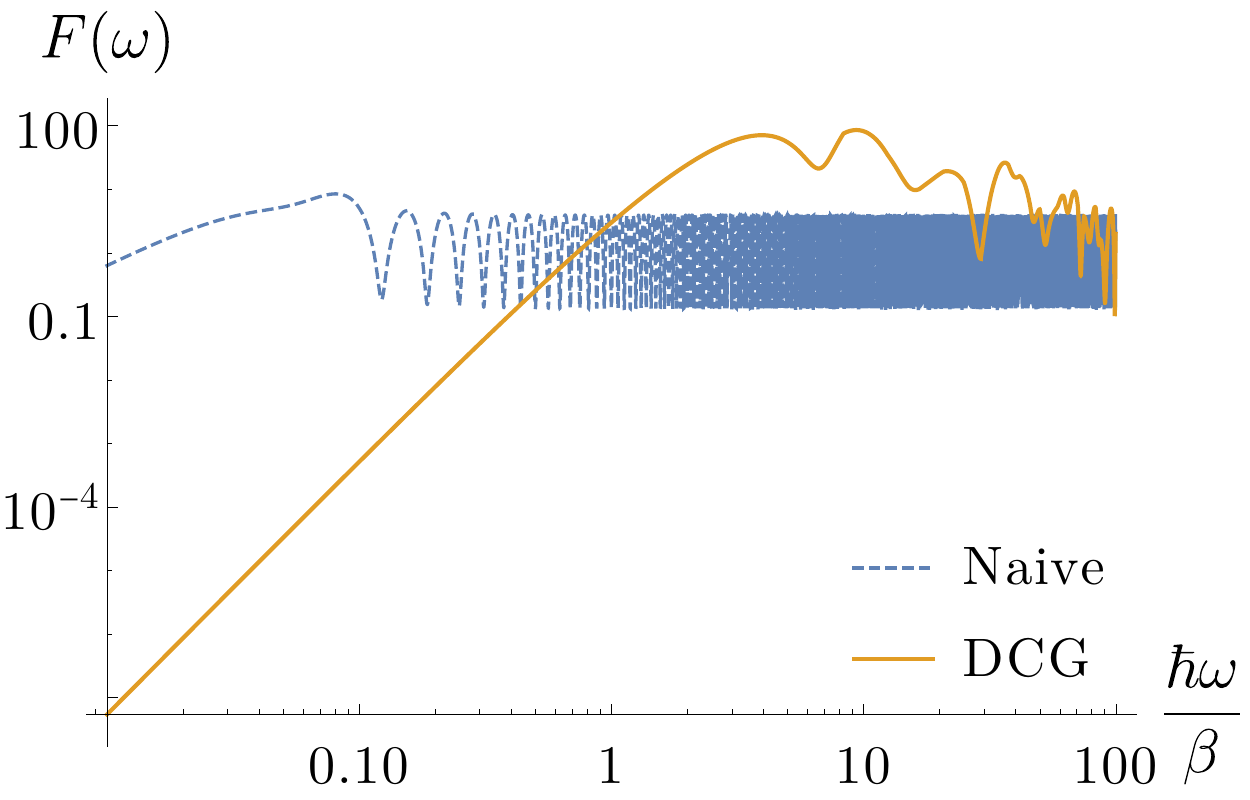}\\
\caption{(Color online) (a) Pulse shape $\Omega_0(t)$ (inset $\Omega(\chi)$) in units of $\beta$ which implements a robust $\theta = 9\pi/5$ rotation around the axis $\boldsymbol n = (\cos\phi,\sin\phi,0)$ with $\phi = \pi/5$. (b) Comparison of the leading order filter functions for the robust gate against a naive implementation using $U_\text{naive}(t)$.}
\label{fig:ex3}
\end{figure}

As an example, we solve for the robust pulse implementing a $U(\theta=9\pi/5, \phi=-\pi/5)$ using the auxiliary equation
\begin{align}
c\partial_t^3 \gamma(t) + \partial_t^6 \gamma(t)=0
\end{align} with the choice $c=300$, $A=\pi/2$ and $n=0$. The pulse shape and the corresponding filter function are shown in Fig.~\ref{fig:ex3}. Compared to the analytical pulse shapes based on the ansatz Eq.~\eqref{eq:ex-ansatz}, we note that this particular auxiliary differential equation leads to numerical solutions which are sharper and take longer time to perform. However, the advantage of the numerical solutions are that they are more flexible in terms of ansatz and allow targeting arbitrary unitaries.

\section{Conclusion}\label{sec:conclusions}
We have shown that it is possible to obtain robust quantum gates using smooth pulses in a completely analytical fashion, which only requires finding a function whose derivative is bounded and satisfies certain local boundary conditions, at initial and final times. This eliminates nonlocal conditions which necessitate a numerical search over auxiliary parameters \cite{Barnes2015,Zeng2019}. Furthermore, we have shown that the problem can also be converted to a set of coupled ODEs, which further eliminates the search for such a bounded function and yields solutions very quickly using standard numerical ODE solvers. Although the presented pulse shapes tend to have narrow peaks, this is due to the simple choices of ansatz and not a fundamental limitation of our approach.

Although our work assumes an $\mathfrak{su}(2)$ algebra, we again emphasize that our results can be applicable to two-qubit scenarios which exhibit that structure. For instance, in $^{28}$Si quantum double dots \cite{Gungordu2018a} or superconducting qubits \cite{Chow2011} with an always-on coupling, the Hamiltonian decouples into two $\mathfrak{su}(2)$ problems, and when the qubits can be addressed separately, each $\mathfrak{su}(2)$ subspace can be controlled separately. Our robust smooth pulses can then be used to suppress exchange noise and eliminate crosstalk while targeting a desired two-qubit unitary.

\begin{acknowledgements}
UG acknowledges helpful discussions with Edwin Barnes and Sophia Economou.
This research was sponsored by the Army Research Office (ARO) and was accomplished under Grant Number W911NF-17-1-0287.
\end{acknowledgements}

\appendix

\bibliography{siqd,extra}

\end{document}